# Case study and analysis of WAN Optimization pre-requirements

Bhargav. Balakrishnan

Chennai, India

bhargav.balakrishnan@gmail.com

*Abstract:* **This paper deals with HOW to analyze the requirements for setting up the WAN Optimizer. The criteria's that needs to be taken into account, the steps involved in the analysis of WAN optimization requirement. These entire analyses will give a complete framework for setting up a WAN optimizer within an organization and the organization will have a clear record on the analysis made before setting up this WAN Optimizer.**

## I. Introduction

In today's world the WAN Network is congested with high traffic which causes a lot of data loss and jitter. In order to bring a solution for this WAN Optimizer (E.g. Riverbed) was introduced over the WAN Network, which will improve the performance of the network to a considerable extent. Even then there are many criteria's that must be fulfilled to attain considerable amount of data transaction provided there is a decrease in the % of data loss over the network. Those criteria's are going to be analysed while discussing about the case study. Here in this article the main focus will be on how to provide an optimized network in places like Banks, Share Markets, large industries that can be IT industry, automobile etc, where data loss and jitter are major criteria's that are taken into account apart from security. Every organization has many applications that are running over the network so the available bandwidth get shared between them but these in turn if not balanced properly would lead to drastic loss in the business, data transactions and the performance of the applications goes down. So how to frame this network provided the investment made by the organization becomes satisfactory.

## I. Criteria's involved in WAN Optimization

Whenever the concept of WAN is taken there are pre requisites that are taken into account like Network Infrastructure, Network Design (Head Office with other Branches), Cable used (Cat 5e, 6 within Organization, Fibre Optic Cable connecting branches, core switches), Model of Router, Model of Switches (Layer 2 or 3), Capacity of the organization, Application that will work on live environment which includes network security like ISA, Forefront, Business solution etc, Servers, Data Base. How the calculations among them are carried out?

i) Network Bandwidth Consumed = TB – TC - (1)

TB – Total Bandwidth provided by ISP
TC – Total bandwidth utilized by the organization during peak time and weekends/non working hours

ii) % of data loss = Data Loss / Total data transmitted *100 - (2)

E.g. Data Loss = 5000Kbps Data Transmitted = 2500 Kbps
5000/2500 * 100 = 200% (Ratio becomes 2:1)

iii) Ratio between % of bandwidth consumed in internal network: % of bandwidth consumed over external network (This provides us the idea on how to manage the network performance during peak hours based upon which the priotization should be done) – (3)

iv) Network Performance = Data transmission over Fibre optic + Data Transmission over Cat 6/5e Cables - (4)

Data Transmission = Total data transmitted – Data loss (Calculate this for Fibre Optic and Cat 6/5e cables) – (4(i))

% of network performance = Total Data transmission/ Total bandwidth provided * 100

v) % of service provided by ISP = (24*7 – Downtime) / (24*7) * 100 – (5)

% of Service provided by ISP: % of overall Output (Expected along with fault tolerance)

vi) Total Consumption (%) = ABC*UBC*BW*DBC*UWBC – (6)

ABC – Application Bandwidth Consumption UBC- User Bandwidth consumption BW- Bandwidth wastage DBC- Data Transaction Bandwidth Consumption UWBC – Unwanted Bandwidth Consumption.

vii) % of N/W used during peak hours : % of N/W usage in non peak hours/ weekend

*Note: - Calculate all of them on an average to get an approximate value on each criterion.*

These are some of the criteria's that will help Infrastructure team to design the network initially and put some of the operations of an application on live to get an idea on the network consumption as a whole. Based upon the overall performance they can configure the WAN Optimizer which will in turn accelerate the performance of the applications which are running live. For E.g. Finacle is now being widely used in the Banking sectors where it will get merged with Oracle Applications, Oracle Database etc. Along with this there are certain commonly used applications like Active Directory service, Exchange Servers, Anti Virus Servers, File Servers, Share Point Servers, Web Server, Blackberry servers etc...So bandwidth consumed by these applications should be studied carefully by the network management, server management and Application management teams. Based upon which the router, WAN Optimizer should be configured to get the best overall output by excluding the % of data Loss. Once the data loss is reduced then the transmission over the network becomes much faster thereby Jitter and data loss can be eliminated to a greater extent and here too it is not 100% as there is always certain amount of data loss and jitter along with reduction of network performance

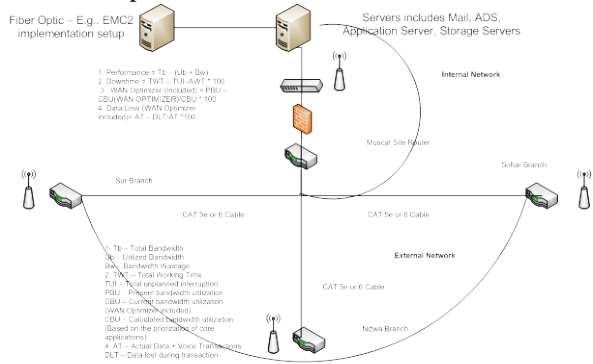

Fig (1)

## II. Case Study of Bank (Large Sectors)

Let us consider a Bank with 200 Employees and 3 Branches
1. Head Office (Muscat) – 75
2. Sur Branch – 25
3. Sohar - 60
4. Nizwa – 40

So the above mentioned applications in the topic criteria's involved in WAN Optimization will be working along with certain other Backup Solution and Data Achieving i.e. Veritas Net Backup and Net Vault. Here the Head Office will mostly have all the servers running the applications which are mentioned above. So how the network is going to be maintained optimized and thereby providing the best results at the end of the financial year. So the network diagram of this is shown below.

According to the above network diagram the branches will depend on the services from the head office. So here the priotization of application will come into the pictures. Since the countries like Oman the telephone exchange outside Muscat is mostly giving a maximum bandwidth of 128Kbps leased line. It is because the exchange can support only that amount of bandwidth. In this case the prioritization will have to come for Finacle and Mail Servers as they are main application that is needed for processing the data transmission with the clients so that by the end of the day the reports can be generated without any issues. The internet access permissions can be given for Managerial level as they need certain other applications through which they need to communicate with the head office like Office communicator, instant messenger etc. In this way the application usage are balanced so that all the data storages, transactions and updation of reports are generated properly for the entire organisation. This shows how well the network is optimized by using a WAN Optimizer. This is what makes the organization satisfied on their investment commonly termed **investment on return.** To achieve this IT Team should design the network in such a way that the above result is obtained constantly. The design of the network should be done with redundancy i.e. MPLS connection that provides 24 hours support from the service provider along with Fault tolerance i.e. if one connection is down the other connection will take over it. But the investment is high for a Medium and small level sectors. This is also taken into account only because the service provided is not interrupted by an unplanned downtime by the ISP. All the above information combined together provides the solution of How to frame a network, prioritize and execute the usage the WAN Optimizer. So entire team of IT and Managers should make a work flow chart on How to bring the best throughput by optimizing the speed of the applications as this section falls under service transition if any change in the application is needed like instead of Finacle if Fin Flex is lighter and more stabilized then that can be used provided all the parameters are satisfied as per the Banks requirement. By making the final consideration of the report the IT Team should make their design, test (Application in Live), execute.

### A. Designing Phase & Test

During the design phase the IT Team will have all the reports that were discussed. Initially the team should start working out on the core applications which is going to be used widely by an organization. From that onwards all the other common application should be taken into account, make a statistics graph showing the approximate bandwidth consumption by each application. An example of statistic graph is shown below.

In that let's take the consumption of Share point portal and Finacle which is used comparatively more, so these applications should be prioritized based upon their importunacy to the organization. These applications will be configured on WAN optimizer so that the performance can be maintained. Even the backup should be taken care in WAN Optimization because that also plays role in storing and retrieving of Data's. So these certain applications should be put on a test before it is launched on a Live. The calculation of the above formulas should be done in test and live then only the IT team will be to get a clear picture on the bandwidth consumption. These calculations can be made as case study for the future up gradation and reference. Based upon this only will tell the IT Team on how successfully they have implemented the infrastructure for an organization.

### B. Live Environment

The organization should always make a periodic test on the performance of all the application over the network. If there is any flaw which is identified should be immediately been analyzed and send the information to the respective team based upon which the change advisory board should decide on an alternate solution e.g. the bandwidth consumption is more during the peak hours. The reason should be placed during the meeting session like 1.The users download unwanted files from the internet 2.The bandwidth utilized by certain application is more mainly it will be the core application 3. The data storage transmission i.e. from the application to the data base 4.Anti Virus and firewall application like ISA, Forefront security etc… These are some of the common issues that come across normally so these needs to justify properly. For justifying these things the team needs to prepare a case study on that application then make a list of testing on all perspectives i.e. for example if is database how the distribute the data's over the network i.e. Network load balancing so that this can be avoided. When these are structured in this way then the WAN Optimizer will play the role of amplifying the performance of the application. In order to get the best throughput of the WAN Optimizer these pre-requisite need to be satisfied. As the performance of all the application should be balanced properly over the network especially for the data base as it will work mostly on clustering.

The test results of all the live servers should be brought towards the management every half year basis as a report to show them how best is the infrastructure s maintained within the organization.

*C. Execution*

This will be final phases after Live Testing of the critical applications. Here the reports of all the testing which were done in the live servers will be verified that the condition satisfies the usage of WAN optimizer and the respective applications for which this needs to be prioritized. How is the prioritization calculated?

1. Prioritization in % = Number of usage * Approx time duration utilized /total no of employee * 100

In approximate time duration time duration of peak hours with + or – extra hours.

2. Downtime in % = Total working hours(Application in WAN optimizer) / Total working hours *100

Downtime is unplanned downtime. Maintenance will be reduced in total working hours itself.

3. Percentage of transactions = Total of transaction done - Total no of lost transaction / Actual number of transaction * 100

So why above calculation are done for fixing a WAN Optimizer? What is need of these testing processes? Hardly the device is going to cost some 10 Lakhs Maximum with installation charges.

The answer for this every inclusion of additional device needs a proper justification and at the time same the investment should have some output in the organizations overall result. These testing are done only to ensure that the critical applications are brought under the umbrella of WAN Optimizer which is going to handle the performance of all the application that is configured under it. After that these testing should be done in order to ensure that there is any change in the speed of processing of the application that they really want to be. This is what is going to produce the final report to the management staffs that WAN optimization is going the enhance the performance of critical applications of an organization by the way the organization can expect this result in the next financial year.

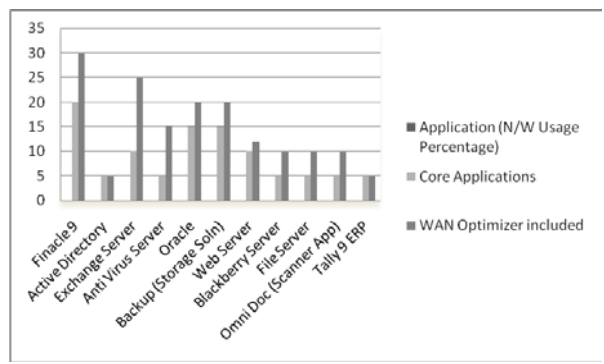

Fig (2)

When the excel sheet of the above graph is opened, there is a prioritization of core applications that is used in Banking Sector. This will be same scenario for other sectors too. So when the WAN Optimizer is tuned in this way then the performance of the core application will be much faster expecially the Daily Backup of the entire Banking process should be properly generated nor it will lead to data loss if there is an interruption in the network performance as the % of data transfer depends the type of cable used especially for the backup Fiber optic cables are used. Especially countries like India the EOD process will be tedious for them this will be quite relieving process rather they need to monitor on each and every

### III. Present Scenario

Many organizations have come across this WAN Optimization. But how far they are analysed and used for the purpose of the organization? How far the end users are satisfied with the performance? Even though the network is optimized with improving the performance of the application it depends on the network they use that is wired or wireless. This also matter a lot in getting the performance of the applications. Then the branch offices which are in out skirt also have to be taken into consideration. All these sectors need to analyze before the live testing. Once these network setups are confirmed then it is a matter of implementation. In present scenario how far these reports are built by the IT team for the purpose of adopting the policies of an organization? How far the organizations adhere to the IT Policies? These are some of the policies that need to be set i.e. group policies, internet restrictions, USB restrictions to get the entire benefit out of this WAN optimization usage.

These needs to be set and testing of entire network need to be performed and according to which the report needs to be generated. This will then go to the change advisory board for the reviewable and this will get attested saying that there is 99% flaws with no fault tolerance in this current network infrastructure. How to attain this level of accuracy matter a lot? This is what the IT Team can't guarantee in many big organizations. To attain this is not a simple task but time is needed to perform these testing to ensure that the network is optimized and the performance of all application will be properly shared in the network ensuring that the % of data loss and jitter is reduced when compared to the record without WAN optimizer.

IV. Process Flow Diagram

The process flow of this implementation should be according to the above Step that i.e. designing and test.

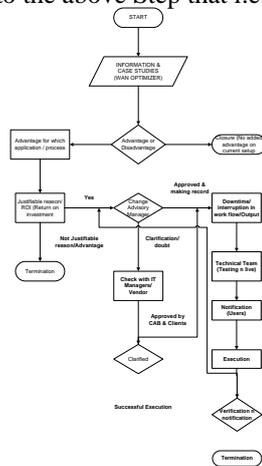

Fig (3)

This workflow will give an idea on **HOW ORGANIZED AN IMPLEMENTATION IS TAKEN**. Since the implementation of WAN Optimization is inter related with other networking devices, the notification should be generated as an Email or SMS (SMS through Blackberry Enterprise server). The clarification needs to clearly documented for future reference if the organization is not been benefitted in anyways by the WAN Optimizer. This work flow involves proper approval at each stage by IT Manager or CAB Member (Involves SLA Manager, Clients, Business relationship manager). So this kind of workflow will help in the executing a task effectively and with proper authentication.

V. Key Advantage on Various WAN Optimizer

In today's world there are various WAN optimizer and each of them have their own features let us some of them like Riverbed, Bluecoat, Juniper and Cisco.

*A. Bluecoat*
Works on the protocol CIFS (common Internet file system) which can significantly reduce the number of round-trips required to satisfy a request, effectively minimizing delays associated with waiting for data retrieval while simultaneously reducing WAN usage

Another major advantage of CIFS protocol is CIFS Protocol Optimization is possible because ProxySG appliances have the ability to terminate user requests as if they were the server. They can then open a separate connection to the server that they can use to intelligently make requests on the client's behalf. The ProxySG appliance, acting as the client, is able to take advantage of the CIFS protocol in a way that the client cannot.

*B. Riverbed*
Riverbed's award-winning wide-area data services (WDS) solutions strategically enable IT to centralize and reduce operational overhead and expense, while improving end-user satisfaction. Riverbed's products have been proven in some of the most demanding and complex networks in the world, with thousands of customers deploying Steelhead WDS solutions. Some Key advantages of Riverbed Steelhead is Speed, Scalability, scalability and Accelerate key enterprise application. Above all there are some major advantages IT Infrastructure consolidation, Serverless branch office, enhanced data protection, enabling the mobile workforce. There are key streaming processing data, application, transport and management which will improve the performance of key enterprise applications by 30%-40% than before i.e. if the current performance is 60% it will be improved to 95% by using the Riverbed Steelhead.

*C.Cisco WAAS*
Cisco has come up the WAN Optimizer known as Wide Area Application Services. This has some major advantages when compared to the above two WAN optimizer in terms of deployment and management. WAAS Central Manager is a web-based central management tool that provides simplified configuration, provisioning, monitoring, fault-management, logging and reporting for up to 2,500 WAEs within a Cisco WAAS topology. Cisco's WAAS requires no modifications to applications, clients, or servers in order to provide acceleration services

Transparency

As mentioned, for many IT organizations the first rule of networking is that you should not do anything that causes the network to break. For example, it should be possible to implement a WAN optimization solution and not break existing functionality such as security, QoS or routing. It should also be possible to implement a WAN optimization solution and still be able to use the existing management and monitoring tools. Given that WAAS does not change the packet headers, it provides a high degree of transparency. That transparency has an impact on TCO as well as the following factor.
• Ease of Deployment and Management
Because it does not modify packet headers, Cisco's WAAS is less likely to break some aspect of IT than are some other solutions. As a result, WAAS is easier to deploy and manage than solutions that break something.

In addition, WAAS supports auto-discovery whereby the solution checks to see if a peer acceleration appliance exists in the packet flow between the source and the destination. This functionality makes deployment easier by eliminating the need to have IT Organizations implement an overlay network.
• Integration
While this was not one of the criteria listed in Table 1, it impacts several of the criteria that were mentioned in that table. The point being that given Cisco's position in the enterprise networking marketplace, the chances are that the environment in which the WAN optimization solutions will be deployed will be a Cisco environment. Cisco is in the best position to ensure that its WAN optimization solution does not impact this environment.
• Performance
Some of the comments that Miercom made about Cisco's WAAS include:
• Under favourable conditions (large WAN latency and highly compressible content) the WAAS V4.0 solution can provide impressive degrees of acceleration.
• The software demonstrated performance parity with these other products, and that in some key metrics provided superior compression, speed and throughput.
• Unlike some of the other products in the market, Cisco's WAAS does not degrade

So each WAN Optimizer manufacturing companies test their device on the aspect of fault tolerance i.e. preventing any disturbance on the current infrastructure setup, in the same way this case studies needs to be carried out the IT team in order to justify the management that the Why they have selected this WAN optimizer and what is the Key advantage of that. This is the main aim behind designing this paper

## VI. Conclusion

Finally concluding this paper on the pre-requirement framework for WAN optimizer with the aim, that the following test cases are prepared with the following calculations mentioned above. This will be mapped with the work-flow diagram. On the whole this will generate a report for setting up the WAN Optimizer efficiently for an organization.